\documentclass[12pt]{article}
\usepackage{amsfonts}
\usepackage{amssymb}
\usepackage{amscd}

\begin{document}

{\bf ON SOME $p$-ADIC SERIES WITH FACTORIALS}

\bigskip

  Branko DRAGOVICH

 Institute of Physics, P. O. Box 57, 11001
Belgrade, Yugoslavia

\bigskip

\bigskip

\bigskip

\section{ Introduction}
 \noindent In the last nine years $p$-adic
numbers have been successfully applied in various branches of
theoretical and mathematical physics (for a review see [1-3]). One
of the objects which we often encounter is a $p$-adic series.

My interest in $p$-adic series is mainly motivated by the
significant role they might play not only in pure but also in
applied mathematics. It was initiated in 1987 [4] by an
observation that divergent perturbative series, which we usually
face up in theoretical and mathematical physics, are $p$-adically
convergent. Namely, such power series have rational coefficients
and may be treated in $R$ as well as in any  $Q_p$. Loosely
speaking, the less convergence in $R$ the more convergence in
$Q_p$, and vice versa.

This opportunity induces a question on possible connection between
convergence in the one number field and summation of divergent
counterpart in the other one. It is natural to expect an answer
within the field of rational numbers $Q$, because $Q$ is a subfield
of $R$ and all $Q_p$.

Recall that the sum of a divergent series depends on the way how
one performs summation. If a series is convergent and has a
definite rational sum  in $Q_p$ for all but a finite number of $p$
then this sum may be attached to the divergent counterpart. In
other words, it seems appropriate to choose the same procedure of
summation for divergent as that one for convergent versions of a
series.

For  often encountered divergent series, the situation is such
that only trivial rational points exist ({\it i.e.} for argument
$x = 0$). However there are divergent power series which are
$p$-adically convergent and have non-trivial (usually one)
rational points. Such series contain factorials in the numerator
of coefficients and they are similar to perturbative expansions in
quantum field theory and string theory.

To illustrate divergence in quantum field theory one can consider the
simple integral for zero-dimensional scalar theory
$$
Z(m, g) = \int_Rd\varphi \exp (-\varphi{m^2\over2}\varphi+g\varphi^3)
$$
which leads to a perturbative series
$$
\sum^\infty_{n=0}{(6n-1)!!\over(2n)!}x^n\ ,
$$
where $x = g^2/m^6\ ,m$ is a mass and $g$ is a coupling constant.
This series, as well as the partition function in perturbation string
theory [5], diverges like $\sum n!x^n$.

On the other hand, to make a direct connection of $p$-adic models
with the corresponding real one it is necessary to have
convergence of series in $R$ and all $Q_p$ within the common
domain of rational numbers. Namely, all results of measurements
belong to $Q$, and comparison of theory and experiment performs
within $Q$. However, the standard power series of theoretical
physics do not satisfy the above condition. For example,
expansions of functions $\exp x\, \sin x\, \cos x\, \sinh x$ and
$\cosh x$ are convergent in $p$-adic case for $\mid x\mid_p<\mid
2\mid_p$. As a consequence, there is no $0\not=x\in Q$ for which
these functions are defined for every $p$. Therefore it is
reasonable to consider small modifications of the standard
expansions which lead to significant enlargement of the region of
convergence in $p$-adic counterparts.

This article is based on the author's papers [6-10]. The wide
class of series considered here is given by expression (2.5).
Enlargement of $p$-adic region of convergence is introduced by a
parameter $q\in Q$ in the denominator of coefficients. The
presence of $q$ in the form (2.3) makes the series (2.5)
convergent everywhere on $Q_p$ for every $p$, and it also does not
affect convergence properties in the real case. This
$q$-modification can be made arbitrarily small in real counterpart
leaving $p$-adic convergence unchanged. For $q = 0$ these series
become in a sense the standard ones. The domain of convergence is
found for all characteristic cases.

The general summation formula (3.1) is derived, which is number
field invariant. When a rational sum is obtained for all but a finite
number of $p$ it may be used for summation of divergent counterparts.
This method of summation of divergent series is called "adelic
summation" and may be used in some rational points.

In the case $q = 0$ we mainly pay attention to the series which
are convergent everywhere on the ring of $p$-adic integers $Z_p$
for every $p$. One can say, the more simple a series is, the more
we go  into details. In particular, the series
$\sum^\infty_{n=0}n!P_k(n)$, where $P_k(n)$ is a polynomial in $n$
and with rational coefficients, is investigated rather widely. It
is obtained a method to find all $P_k(n)$ which yield rational
sums. The connection between uniqueness of the pair of integers
$(u_k ,v_k)$ in $\sum n!(n^k+u_k) = v_k$ and possible
non-rationality of $\sum n!$ is pointed out. Although there is not
yet a proof that $\sum n!$ is not a rational number it seems very
likely that it is true. A proof of this conjecture might be very
significant.

Since the concept of adeles (Platonov and Rapinchuk [11]) enables
to consider properties of $Q$ simultaneously from real  and
$p$-adic points of view, in Section 5 one investigates some adelic
aspects of our series. It is shown that one can make an adelic
sequence of some series at rational points.

In the Appendix we give some examples of characteristic series
with a rational sum to illustrate their non-triviality and
diversity .

All necessary $p$-adic analysis needed for investigation of these
series can be found in an excellent book of Schikhof [12].

\bigskip

\section{ Convergence}

\noindent If we have a power series
$$
\sum^\infty_{n=0}A_nx^n\ ,\eqno(2. 1)
$$
where $A_n\in Q$, it can be treated as a $p$-adic series $(x\in
Q_p)$ as well as a real one $(x\in R)$. Recall that, in the
$p$-adic case, a necessary and sufficient condition for a
convergence of (2. 1) is
$$
\mid A_nx^n\mid_p\to 0\quad\hbox{as}\quad n\to\infty\ .\eqno(2. 2)
$$
Let
$$
I^{(q)}_{\mu n+\nu} = {\big((\mu n+\nu)!\big)^{\mu n+\nu}\over q+
\big((\mu n+\nu)!\big)^{\mu n+\nu}}\ ,\eqno(2. 3)
$$
where $\mu\in Z_+ = \{1, 2, \cdots\}\ ,\quad\nu\in Z_0 = \{0, 1,
2,\cdots\}$ and $q$ is a nonnegative rational number $(0\leq q\in
Q)$. Let also
$$
P_k(n) = C_kn^k+\cdots +C_0\eqno(2. 4)
$$
be a polynomial in $n\in Z_+$ of degree $k$ with the coefficients
$C_0, \cdots , C_k\in Q$.

\bigskip

\noindent
 {\bf Proposition 1 \ \ }
 The power series
$$
\sum^\infty_{n=0}\varepsilon^nI^{(q)}_{\mu n+\nu}\prod^I_{i=1}\big
((\alpha_in+\beta_i)!\big)^{\lambda_i}P_k(n)x^{\mu n+\nu}\ ,\eqno(2. 5)
$$
where $\varepsilon = +1 \ \hbox{or} \ -1\ , I^{(q)}_{\mu n+\nu}$
is defined by (2. 3) with $q\not=0\ , \alpha_i\ , I\in Z_+\ ,
\beta_i\in Z_0\ , \lambda_i\in Z\ ,$ and $P_k(n)$ is given by (2.
4), is $p$-adically convergent for all $x\in Q_p$ and for any
prime number $p$.

\bigskip

\noindent {\it Proof: \ \ } The general term of (2.5) has a
$p$-adic norm
$$
\mid I^{(q)}_{\mu n+\nu}\mid_p\prod^I_{i=1}\mid
(\alpha_in+\beta_i)!\mid^{\lambda_i}_p\mid P_k(n)\mid_p\mid x\mid^
{\mu n+\nu}_p\ ,\eqno(2. 6)
$$
where $\mid P_k(n)\mid_p\leq \ \mathop{{\rm max}}\limits_{0\leq j\leq k}\mid
C_j\mid_p\ ,$ and for a large enough $n$
$$
\mid I^{(q)}_{\mu n+\nu}\mid_p = \mid q\mid^{-1}_p\mid (\mu
n+\nu)!\mid_p^{\mu n+\nu}\eqno(2. 7)
$$
as a consequence of the strong triangle inequality for $p$-adic
norm. Recall that
$$
\mid m!\mid_p = p^{-{m-S_{m}\over p-1}}\ ,\eqno(2. 8)
$$
where $S_m$ is the sum of digits in the canonical expansion of a
positive integer $m$ over $p$. Thus, for a large enough $n$ (2.6)
behaves like
$$
\bigg\{p^{-{\mu\over p-1}n}p^{-{1\over (p-1)\mu}\sum^{I}_{i=1}\alpha_{i}
\lambda_{i}}\mid x\mid_p\bigg\}^{\mu n}\ .\eqno(2. 9)
$$
For any $x\in Q_p$ and for the above range of parameters the
expression (2.9) tends to zero as $n\to\infty$. \bigskip

 \noindent
{\bf Proposition 2 \ \ } The power series (2.5) for the same
parameters as in Proposition 1, but $q = 0$, is $p$-adically
convergent in the domain
$$
\mid x\mid_p<p^{{1\over (p-1)\mu}\sum^{I}_{i=1}\alpha_{i}
\lambda_{i}}\ .\eqno(2. 10)
$$

\bigskip

\noindent {\it Proof: \ \ } Since $I^{(0)}_{\mu n+\nu}\equiv 1$,
instead of (2.9) one obtains
$$
\bigg\{p^{-{1\over (p-1)\mu}\sum^{I}_{i=1}\alpha_{i}
\lambda_{i}}\mid x\mid_p\bigg\}^{\mu n}\eqno(2. 11)
$$
which tends to zero as $n\to\infty$ iff $\mid x\mid_p$ satisfies (2.10).

In order to have $p$-adic convergence for any $x\in Q$ at all but
a finite number of $p$, we will be mainly interested in the cases
when
$$
\sum^I_{i=1}\alpha_i\lambda_i\geq\mu\eqno(2. 12)
$$
in (2.11).

While in the $p$-adic case factors $I^{(q)}_{\mu n+\nu}$, for
$q\not=0$, serve to extend the domain of convergence, in the real
one they do not play a such role. Namely, since in the real case
$I^{(q)}_ {\mu n+\nu}\to1$ as $n\to\infty$, these factors do not
influence a change of the domain of convergence of series (2. 5)
for various values of the parameter $q$. However, other parameters
are more or less important and convergence in detail may be
determined using the d'Alembert criterion.

In particular, the following two simple classes of (2. 5) deserve to
be noted:
$$
E^{\varepsilon, q}_{\mu,\nu}(x) =
\sum^\infty_{n=0}\varepsilon^nI^{(q)}_{\mu n+\nu}{x^{\mu
n+\nu}\over(\mu n+\nu)!}\ ,\eqno(2. 13)
$$
which is everywhere convergent in $R$ and all $Q_p$ if $q\not=0$, and
for $\mid x\mid_p<\mid 2\mid_p$ if $q=0$;
$$
F^{\varepsilon, q}_{\mu,\nu}(x) =
\sum^\infty_{n=0}\varepsilon^nI^{(q)}_{\mu n+\nu}(\mu
n+\nu)!x^{\mu n+\nu}\ ,\eqno(2. 14)
$$
which is everywhere divergent in $R$, everywhere $p$-adic
convergent if $q\not=0$, and $p$-adic convergent for $\mid
x\mid_p\leq1$ if $q=0$. Series of the form (2.13) contain
$q$-modification of the expansions for the well-known functions
like exponential $(\varepsilon = 1 ,\mu = 1 ,\nu = 0), \ {\rm
cosine} \ (\varepsilon = -1 ,\mu = 2 ,\nu = 0)$ and ${\rm sine} \
(\varepsilon = -1 ,\mu = 2 ,\nu = 1)$. Expansion (2.14) has
factors $(\mu n+\nu)!$ which are inverse to (2.13).

\section{ Summation}

 \noindent Starting from the
series (2. 3) and owing to the factorization of expressions with
factorials one can obtain a summation formula for a wide class of
series.

Let $(m+1)_\mu = (m+1)(m+2)\cdots(m+\mu)\ .$ \bigskip

 \noindent
{\bf Proposition 3\ \ } The summation formula
$$
\sum^\infty_{n=0}\varepsilon^n\bigg((\mu n+\nu)!\bigg)^{\mu
n+\nu}\prod^I_{i=1}
\bigg((\alpha_in+\beta_i)!\bigg)^{\lambda_{i}}$$  $$ \times
\bigg\{{\bigg((\mu n+\nu)!\bigg)^\mu(\mu
n+\nu+1)^{\mu(n+1)+\nu}_\mu\over
q+\bigg[\bigg((\mu(n+1)+\nu)!\bigg]^{\mu(n+1)+\nu}} \prod^I_{i=1}
(\alpha_in+\beta_i+1)^{\lambda_{i}}_{\alpha_{i}}P_k(n+1)x^\mu  $$
$$
 -\varepsilon{1\over q+\bigg((\mu n+\nu)!\bigg)^{\mu
n+\nu}}P_k(n)\bigg\} x^{\mu n+\nu} $$  $$ = -\varepsilon{(\nu
!)^\nu\over q+(\nu
!)^\nu}\prod^I_{i=1}(\beta_i!)^{\lambda_{i}}P_k(0)x^\nu \eqno(3.1)
$$
has a place in the region of parameters and variable $x$ which are determined
by convergence of the series
(2.5).

\bigskip

\noindent {\it Proof:\ \ } Taking into account the identity
$$
\bigg[\bigg((\mu(n+1)+\nu)!\bigg]^{\mu(n+1)+\nu} = \bigg((\mu n+\nu)!\bigg)
^{\mu n+\nu}\bigg((\mu n+\nu)!\bigg)^\mu(\mu n+\nu+1)^{\mu
(n+1)+\nu}_\mu \eqno(3. 2)
$$
the LHS of expression (3.1) may be rewritten as
$$
\sum^\infty_{n=1}\varepsilon^{n-1}I^{(q)}_{\mu n+\nu}
\prod^I_{i=1}
\bigg((\alpha_in+\beta_i)!\bigg)^{\lambda_{i}}P_k(n)x^{\mu n+\nu}
$$ $$- \sum^\infty_{n=0}\varepsilon^{n-1}I^{(q)}_{\mu n+\nu}
\prod^I_{i=1}
\bigg((\alpha_in+\beta_i)!\bigg)^{\lambda_{i}}P_k(n)x^{\mu
n+\nu}$$ $$ =\varepsilon I^{(q)}_\nu\prod^I_{i=1}
(\beta_i!)^{\lambda_{i}}P_k(0)x^\nu . \eqno(3. 3)
$$

Although the formula (3.1) is based on mutual cancellation in
pairs of terms in the LHS of (3.3), leaving only the first one, it
is very useful and yields highly non-trivial results.

The summation in (3.1) does not depend on the number field. It is
worth noting that for any $x\in Q$ the sum is a definite rational
number given by the RHS and this result holds in all $Q_p$ if
$q\not=0$.

In the real case the LHS of (3.1) can be divergent. If so, a sum
depends on the way of summation. In such case, among all
reasonable ways of summation it seems that the number field
invariant one is the most natural. \bigskip

 \noindent {\bf Definition\ } (Adelic summation) Let a series be
 divergent in the real case
and convergent in $Q_p$ for all but a finite number of $p$. Let
such series allows a number field invariant summation with
rational sum for some variable $x\in Q$ in the domain of
convergence. Extrapolation of the number field invariant summation
to the divergent counterparts we will call adelic summation.

\section{ Case  q = 0}

 \noindent The results
obtained in the previous sections are mainly related to the series
(2.5) with $q\not=0$, where two particular cases are noted: (2.13)
and (2.14). Here we will consider the case $q=0$, {\it i.e.} we
will investigate some aspects of
$$
\sum^\infty_{n=0}\varepsilon^n\prod^I_{i=1}\bigg((\alpha_in+\beta_i)!\bigg)^
{\lambda_{i}}P_k(n)x^{\mu n+\nu}\ .\eqno(4. 1)
$$
Recall that the domain of convergence of (4.1) is already derived
and given by (2.10). The corresponding summation formula follows
from (3.1) and it reads:
$$
\sum^\infty_{n=0}\varepsilon^n\prod^I_{i=1}\bigg((\alpha_in+\beta_i)
!\bigg)^{\lambda_{i}}\bigg\{\prod^I_{i=1}(\alpha_in+\beta_i+1)^
{\lambda_{i}}_{\alpha_{i}}P_k(n+1)x^\mu $$ $$-\varepsilon
P_k(n)\bigg\}x^{\mu n+\nu} = -\varepsilon\prod^I_{n=1}
(\beta_i!)^{\lambda_{i}}P_k(0)x^\nu\ .  \eqno(4. 2)
$$
In some pure theoretical, as well as practical problems a
knowledge on existence of non-trivial rational points may be very
important. \bigskip

 \noindent {\bf Proposition 4\ \ } The series (4.1)
has a rational sum for some $x\in Q$ which satisfies (2.10), if
there exists a polynomial $A_\eta(n)$ such that
$$
P_k(n) = \prod^I_{i=1}(\alpha_in+\beta_i+1)^{\lambda_{i}}_{\alpha_{i}}
x^\mu A_\eta(n+1)-\varepsilon A_\eta(n)\ .\eqno(4. 3)
$$
\bigskip

\noindent {\it Proof: \ \ } Let there exists a polynomial
$A_\eta(n)$ with rational coefficients which satisfies equation
(4.3), then according to formula (4.2) the rational sum of (4.1)
does exist and the sum is
$$
{\cal S} = -\varepsilon\prod^I_{i=1}(\beta_i!)^{\lambda_{i}}A_\eta(0)
x^\nu\ .\eqno(4. 4)
$$

Note that Proposition 4 defines a sufficient condition. So far, on
a necessary condition one can only conjecture. It is clear that
all series (4.1) may not have a non-trivial rational point ({\it
i.e.} for $x\not=0$).

\bigskip

 \noindent{\bf  Proposition 5\ \ } For a given
$x\in Q$ in the series (4.1) one can always find a polynomial
$P_k(n)$ so that the sum is a rational number. The degree of
$P_k(n)$ is $k = {\hbox {max}}\{\sum^I_{i=1}\alpha_i\lambda_i
+\eta\ ,\eta\}$.

\bigskip

\noindent {\it Proof: \ \ } According to (4.3) for a given $x\in
Q$ there is a polynomial $P_k(n)$ which depends on a particular
choice of a polynomial $A_\eta(n)$. The degree of $P_k(n)$ also
follows from (4.3).

If $\sum^I_{i=1}\alpha_i\lambda_i\geq1$, then $k = \sum^I_{i=1}\alpha_i
\lambda_i+\eta$. Condition (2.12) belongs to this case.

The simplest factorial form of (4.1) with $P_k(n)$ is
$$
\sum^\infty_{n=0}n!P_k(n)\ ,\eqno(4. 5)
$$
which is divergent in $R$ and convergent in all $Q_p$. The
corresponding summation formula is
$$
\sum^\infty_{n=0}n!\big[(n+1)A_\eta(n+1)-A_\eta(n)\big] = -A_\eta(0)\
,\eqno(4. 6)
$$
where $A_\eta(n) = a_\eta n^\eta+a_{\eta-1}n^{\eta-1}+\cdots+a_0$,
with $a_\eta,\cdots,a_0\in Q$.

It is obvious that all possible polynomials $A_\eta(n)$ generate the
corresponding polynomials $P_k(n)\ ,k = \eta+1\geq1$, which allow
rational sums $-A_\eta(0)$. Searching of all possible $P_k(n)$ may be
reduced to
$$
\sum^\infty_{n=0}n!(n^k+u_k) = v_k\ ,(k\geq1)\ ,\eqno(4. 7)
$$
where $u_k, v_k\in Q$ have to be determined, {\it i.e.} one has to
solve equation
$$
(n+1)A_{k-1}(n+1)-A_{k-1}(n) = n^k+u_k\ .\eqno(4. 8)
$$

To Eq. (4.7) corresponds a system of $k+1$ linear equations with
$k+1$ unknowns $(a_0, a_1,\cdots,a_{k-1}, u_k)$, which has always
a solution. Note that $v_k = -A_{k-1}(0)$. The first five of pairs
$(u_k, v_k)$ are: $(u_1, v_1) = (0, -1)\ ,(u_2, v_2) = (1, 1)\
,(u_3, v_3) = (-1, 1)\ , (u_4, v_4) = (-2, -5)\ ,(u_5, v_5) = (9,
5)$.

\bigskip \noindent {\bf Proposition 6\ \ } If pairs of rational
numbers $(u_k, v_k)$ are not unique for a given $k\geq1$, then
$\sum^\infty_{n=0}n!$ is a rational number in a $Q_p$.

\bigskip

\noindent { \it Proof: \ \ } Suppose that in addition to $(u_k,
v_k)$ exists $(u_k^\prime, v^\prime_k)\not=(u_k, v_k)$. Then
$\sum^\infty_{n=0}n! = (v_k-v_k^\prime)/(u_k-u_k^\prime)$.

\bigskip

\noindent {\bf Proposition 7\ \ } If $\sum^\infty_{n=0}n!$ is a
rational number then $\sum^\infty_{n=0}n!n^k$, for any $k\in Z_+$,
are also rational numbers.

\bigskip

\noindent { \it Proof:\ \ } It follows from (4.7). \bigskip

There is not yet an exact proof that $\sum^\infty_{n=0}n!$ is not
a rational number (Schikhof [12]). But it seems reasonable to
suppose that it is not a rational number. If so, then pairs $(u_k,
v_k)$ are unique, and existence of a polynomial $A_\eta(n)$ is not
only sufficient but also a necessary condition for rational
summation of (4.5). Thus the general form of (4.5) with rational
sums is
$$
\sum^\infty_{n=0}n!(C_kn^k+C_{k-1}n^{k-1}+\cdots+C_0) = D_k\ ,
$$
where $C_0 = \sum^k_{j=1}C_ju_j\ , D_k = \sum^k_{j=1}C_jv_j\ ,$
and $C_1\cdots ,C_k\in Q$.

Note that the above consideration performed for (4.5) can be done for
$$
\sum^\infty_{n=0}(-1)^nn!P_k(n)
$$
with analogous conclusions. \bigskip

 \section{ Adelic Aspects}

  \noindent Recall (Platonov and
Rapinchuk [11]) that an adele is an infinite sequence
$$
x = (x_\infty, x_2,\cdots, x_p,\cdots)\ ,\eqno(5. 1)
$$
where $x_\infty\in R,\, a_p\in Q_p$ with a restriction that
$a_p\in Z_p = \{t:\mid t\mid_p\leq1\}$ for all but a finite number
of $p$. The set of all adeles is a ring under componentwise
addition and componentwise multiplication. The space of adeles
${\cal A}$ may be presented as
$$
{\cal A} = \bigcup \limits_{S}A(S)\ , \, \, \,  A(S) =
R\times\prod_{p\in S}Q_p \times\prod_{p\not\in S}Z_p\ ,\eqno(5. 2)
$$
where $S$ is a set of finite number of primes $p$. ${\cal A}$, as
a topological space, has a basis of open sets which are of the
form $W_\infty \times\prod_{p\in S}W_p\times\prod_{p\not\in
S}Z_p$, where $W_\infty$ and $W_p$ are open sets in $R$ and $Q_p$,
respectively.

Note that ${\cal A}$ is an instrument which enables a simultaneous
treatment of all completions of $Q$. The field $Q$ can be embedded
into ${\cal A}$ by mapping $x\to(x, x, \cdots, x, \cdots)$, where
$x\in Q$.

\bigskip \noindent {\bf Proposition 8\ \ } Let one has a sequence
$$
E^\varepsilon_{\mu,\nu}(x) = \bigg(E^{\varepsilon,0}_{\mu,\nu}(x_\infty),
E^{\varepsilon,2^{-s}}_{\mu,\nu}(x_2),\cdots,E^{\varepsilon,p^{-s}}_{\mu,\nu}
(x_p),\cdots\bigg)\ ,\eqno(5. 3)
$$
where $E^{\varepsilon,0}_{\mu,\nu}(x_\infty)$ is a real and
$E^{\varepsilon,p^ {-s}}_{\mu,\nu}(x_p)$ is a $p$-adic series
defined by (2.13), and $s\in Z_+$. When $x = (x_\infty,
x_2,\cdots, x_p,\cdots)$ is an adele, $E^\varepsilon_{\mu,\nu}(x)$
is also an adele.

\bigskip

\noindent {\it Proof:\ \ } For any $x_\infty\in R$, the series
$E^{\varepsilon,0}_{\mu,\nu} (x_\infty)$ is well defined in real
case. The general term of the $p$-adic series (2.13) for $q =
p^{-s}$ is
$$
\varepsilon^n{p^s\big((\mu n+\nu)!\big)^{\mu n+\nu-1}\over1+p^s
\big((\mu n+\nu)!\big)^{\mu n+\nu}}x_p^{\mu n+\nu}\eqno(5. 4)
$$
Since $\mid1+p^s\big((\mu n+\nu)!\big)^{\mu n+\nu}\mid_p = 1$, the
$p$-adic norm of (5.4) is
$$
{1\over p^s}\mid\big((\mu n+\nu)!\big)^{\mu n+\nu-1}\mid_p\mid x_p\mid_p^
{\mu n+\nu}\ ,\eqno(5. 5)
$$
which can be larger than 1 only for a finite number of $p$.

Thus (5.3) is an adelic sequence of series. It is clear that in
(5.3) one can take $x_\infty = x_2 = \cdots =x_p = \cdots = x$,
where $x\in Q$.

\bigskip \noindent { \bf Proposition 9\ \ } The sequence of series
$$
\bigg(H^{\varepsilon,q}_{\mu,\nu}(x),
H^{\varepsilon,q}_{\mu,\nu}(x),\cdots,H^{\varepsilon,q}_{\mu,\nu}
(x),\cdots\bigg)\ ,\eqno(5. 6)
$$
where
$$
H^{\varepsilon,q}_{\mu,\nu}(x) = \sum^\infty_{n=0}\big((\mu
n+\nu)!\big) ^{\mu n+\nu-1} $$  $$ \times\bigg\{{\big((\mu
n+\nu)!\big)^\mu(\mu n+\nu+1)^{\mu(n+1)+\nu-1}_\mu\over
q+\bigg[\big((\mu(n+1)+\nu\big)!\bigg]^{\mu(n+1)+\nu}}x^\mu-
{1\over q+\big((\mu n+\nu)!\big)^{\mu n+\nu}}\bigg\}x^{\mu n+\nu}\
,  \eqno(5. 7)
$$
is an adele of series if $x\in Q$.

\bigskip

\noindent { \it Proof:\ \ } Note that
$H^{\varepsilon,q}_{\mu,\nu}(x)$ is a particular case of the LHS
in (3.1) induced by (2.13), which is convergent for every $x\in R$
and every $x\in Q_p$. It has a sum
$$
{\cal S} = -{(\nu!)^{\nu-1}\over q+(\nu!)^\nu}x^\nu\ ,\eqno(5. 8)
$$
which for $x\in Q$ is a $p$-adic integer for all but a finite
number of $p$.

In virtue of (4.6) one can construct adeles from
$\sum^\infty_{n=0}n!P_k(n)$ with $P_k(n) =
(n+1)A_{k-1}(n+1)-A_{k-1}(n)$ where $A_{k-1}(n)\ ,(k\geq1)$, are
arbitrary polynomials with rational coefficients. The
corresponding sum $-A_{k-1}(0)$, which is valid for all $p$-adic
cases, may be also attributed to the real case by method of adelic
summation of divergent series. \bigskip

 \section{ Appendix}
 \noindent Here we give some particular examples which illustrate
various forms of $p$-adic series with rational sums contained in
the preceding sections of this article. These series are
convergent in all $Q_p$ and have number field independent rational
sum. The first series is also convergent in $R$, but all other are
not. Some other examples can be found in the author's paper [7].
\vskip0.5cm \noindent

$$
\sum^\infty_{n=0}(-1)^n\bigg[{\big((n+1)!\big)^n\over q+\big((n+1)
!\big)^{n+1}}+{(n!)^{n-1}\over q+(n!)^n}\bigg] = {1\over q+1}\
.\quad0\leq q\in Q\ .\eqno(A1)
$$

$$
\sum^\infty_{n=0}n!\big[C_5n^5+C_4n^4+C_3n^3+C_2n^2+C_1n+
(9C_5-2C_4-C_3+C_2)\big] $$ $$= 5C_5-5C_4+C_3+C_2-C_1\ .\quad
C_1,\cdots,C_5\in Q\ . \eqno(A2)
$$

$$
\sum^\infty_{n=0}(n+\beta)!\big[C_2n^2+C_1n-C_2\beta^2
+C_1\beta+C_2\big]  $$ $$= \beta !\big[C_2(\beta+1)-C_1\big]\
.\quad C_1, C_2\in Q\ .   \eqno(A3)
$$

$$
\sum^\infty_{n=0}(2n+\beta)!\big[4n^2+2(2\beta +3)n+\beta^2+3\beta
+1\big] = -\beta !\ .\eqno(A4)
$$

$$
\sum^\infty_{n=0}(2n+\beta)!\big[8n^3-2(3\beta^2 +9\beta
+8)n-2\beta^3-9\beta^2 -11\beta -1\big] $$  $$ = \beta !(2\beta
+5)\ .   \eqno(A5)
$$

$$
\sum^\infty_{n=0}(2n+\beta)!{4n^2+2(2\beta+3)n+\beta(\beta
+3)\over2^n} = -\beta !2\ . \eqno(A6)
$$

$$
\sum^\infty_{n=0}(2n+\beta)!{8n^3-6(\beta^2+3\beta
+3)n-2\beta^3-9\beta^2 -9\beta+4\over2^n} $$  $$ = \beta !2(2\beta
+5)\ .   \eqno(A7)
$$

$$
\sum^\infty_{n=0}\big((n+\beta)!\big)^2\big[n^2+2(\beta+1)n+\beta^2
+2\beta \big] = -(\beta !)^2\ .\eqno(A8)
$$

$$
\sum^\infty_{n=0}\big((n+\beta)!\big)^2\big[n^3-(3\beta^2
+6\beta+4)n-2(\beta+1)^3+2\beta+3\big] $$ $$ = (\beta !)^2(2\beta
+3)\ . \eqno(A9)
$$

$$
\sum^\infty_{n=0}\big((n+\beta)!\big)^2{n^2+2(\beta
+1)n+\beta^2+2\beta -1\over2^n} = -2(\beta !)^2\ .\eqno(A10)
$$

$$
\sum^\infty_{n=0}\big((n+\beta)!\big)^2{n^3-(3\beta^2+6\beta
+5)n-2(\beta+1)^3+2(2\beta +3)\over2^{n+1}} $$ $$ = (\beta
!)^2(2\beta+3)\ . \eqno(A11)
$$

$$
\sum^\infty_{n=0}(n+\beta_1)!(n+\beta_2)!\big[n^2+(\beta_1+\beta_2+2)n+(\beta_1
+1)(\beta_2+1)-1\big] $$ $$ = -(\beta_1)!(\beta_2)!\ . \eqno(A12)
$$

$$
\sum^\infty_{n=0}(n+\beta_1)!(n+\beta_2)!\big[n^3-(\beta_1^2+\beta^2_2+
\beta_1\beta_2+3\beta_1+3\beta_2+4)n $$ $$
-(\beta_1+\beta_2+2)(\beta_1\beta_2+\beta_1+\beta_2)+1\big] =
(\beta_1)!(\beta_2)!(\beta_1+\beta_2+3)\ . \eqno(A13)
$$

$$
\sum^\infty_{n=0}(-1)^n(n+\beta_1)!(n+\beta_2)!\big[n^2+(\beta_1+\beta_2+2)n+
(\beta_1+1)(\beta_2+1)+1\big] $$ $$ = (\beta_1)!(\beta_2)!\ .
 \eqno(A14)
$$

$$
\sum^\infty_{n=0}(-1)^n(n+\beta_1)!(n+\beta_2)!\big[n^3-(\beta_1^2+\beta_2^2
+\beta_1\beta_2+3\beta_1+3\beta_2+2)n  $$ $$
-(\beta_1+\beta_2+2)(\beta_1\beta_2+\beta_1+\beta_2+2)-1\big] =
-(\beta_1)!(\beta_2)!(\beta_1+\beta_2+3)\ .  \eqno(A15)
$$
$$
\sum^\infty_{n=0}\varepsilon^n\prod^I_{i=1}(\alpha_in+\beta_i)!\big[\prod^I_{i=1
}(\alpha_in+\beta_i+1)_{\alpha_{i}}(n+1)^k-\varepsilon n^k\big] =
0\ .\eqno(A16)
$$
\vfill\eject


\begin{thebibliography}{99}

\bibitem {[1]} L. Brekke and P. G. O. Freund, $p$-Adic Numbers in Physics, {\it
Phys. Rep.}, {\bf 233}: 1-66 (1993).

\bibitem{[2]} V. S. Vladimirov, I. V. Volovich and E. I. Zelenov,
{\it $p$-Adic Analysis and Mathematical Physics}, World
Scientific, Singapore, 1994.

\bibitem{[3]} A. Khrennikov, {\it $p$-Adic Valued Distributions in
Mathematical Physics}, Kluwer Academic Publishers, Dordrecht,
1994.

\bibitem{[4]} I. Ya. Aref'eva, B. Dragovich and I. V. Volovich, On
the $p$-Adic Summability of the Anharmonic Oscillator, {\it Phys.
Lett.}, {\bf B \ 200}: 512-514 (1988).

\bibitem{[5]} D. J. Gross and V. PeriwalI, String Perturbation
Theory Diverges, {\it Phys. Rev. Lett.}, {\bf 60}: 2105-2108
(1988).

\bibitem{[6]} B. Dragovich, $p$-Adic Perturbation Series and Adelic
Summability, {\it Phys. Lett.}, {\bf B \ 256}: 392-396 (1991).

\bibitem{[7]} B. Dragovich, On $p$-Adic Aspects of Some Perturbation
Series, {\it Teor. Mat. Fiz.}, {\bf 93}: 211-218 (1992).

\bibitem{[8]} B.  Dragovich, Power Series Everywhere Convergent on
$R$ and All $Q_p$, {\it J. Math. Phys.}, {\bf 34}: 1143-1148
(1993); math-ph/0402037.

\bibitem{[9]} B. Dragovich, Rational Summation of $p$-Adic Series,
{\it Teor. Mat. Fiz.}, {\bf 100}: 342-353 (1994).

\bibitem{[10]} B. Dragovich, On $p$-Adic Series in Mathematical
Physics, {\it Proc. V. A. Steklov Inst. Math.}, {\bf 203}: 255-270
(1994).

\bibitem{[11]} V. P. Platonov and A. S. Rapinchuk, {\it Algebraic
Groups and Number Theory}, Nauka, Moscow, 1990.

\bibitem{[12]} W. H. Schikhof, {\it Ultrametric Calculus: An
Introduction to $p$-Adic Analysis}, Cambridge University Press,
Cambridge, 1984.


\end{thebibliography}
\end{document}